\documentclass[a4paper]{article}
\usepackage{graphicx}
\usepackage{subfig}

\title{Locality and Translations in Braided Ribbon Networks}
\author{Jonathan Hackett\thanks{Email address:
jhackett@perimeterinstitute.ca}\\
Perimeter Institute for Theoretical Physics,\\
31 Caroline St. N., Waterloo, Ontario N2L 2Y5 , Canada, and \\
Department of Physics, University of Waterloo,\\
Waterloo, Ontario N2L 3G1, Canada\\ }

\begin{document}

\maketitle

\begin{abstract}

An overview of microlocality in braided ribbon networks is
presented.  Following this, a series of definitions are presented to
explore the concept of microlocality and the topology of ribbon
networks.  Isolated substructure of ribbon networks are introduced,
and a theorem is proven that allows them to be relocated.  This is
followed by a demonstration of microlocal translations.
Additionally, an investigation into macrolocality and the
implications of invariants in braided ribbon networks are presented.

\end{abstract}

\section{Introduction}

In the last century, there have been repeated discoveries of
underlying structure.  Moving from macroscopic objects, to atoms, to
components of the nuclei, to quarks, it has been demonstrated
repeatedly that the differences between supposedly fundamental
particles are, in fact, merely consequences of the composite
structure of underlying reality. It only seems a natural progression
that such an approach of looking for underlying structure be used to
explain the particles of the standard model. Attempts towards this
end, dubbed preon models, met with many obstacles, but still there
was something deeper that presented itself as a
difficulty.\cite{Pati:1974yy,
Terazawa:1980hh,Harari:1979gi,Shupe:1979fv,Harari:1980ez} The
difficulty is that, as such a process does not have an end, we can
continue to suppose that below the currently understood structure is
another set of \textit{more} fundamental particles. This idea
quickly becomes unappealing at a philosophical level, or even a
practical level, as the question then becomes ``What could make it
end?''.  The idea that the preons would be as fundamental as
possible, such as those in \cite{Bilson-Thompson:2005bz}, provides a
way of achieving the desired end. One way to achieve this end is to
suggest that the preons be composed of structure within space-time.
This suggestion gains further appeal by its convergence with recent
approaches to quantum gravity.

Such a preon model was recently proposed in
\cite{Bilson-Thompson:2005bz} and then extended to the idea of
quantum gravity in \cite{Bilson-Thompson:2006yc}.  The idea of
having a composite model of particle physics that is based upon
topology in quantum gravity is appealing.  The most obvious basis
for its appeal is that such a theory may be viewed as progress
towards a grand unified theory.

I shall investigate some features of this model and the topology of
the structures that it introduces.  Based on this I will discuss the
evolution algebra of this theory, and demonstrate that translations
of the large scale structures are a feature of the theory.

\section{Braided Ribbon Networks}

\begin{figure}[!h]
  \begin{center}
    \includegraphics[scale=0.4]{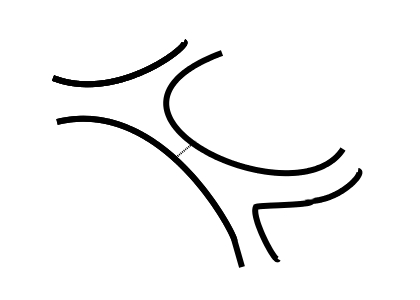}
  \end{center}
\caption{Two trinions with a scored ribbon} \label{figure1}
\end{figure}

The theory of braided ribbon networks \cite{Bilson-Thompson:2006yc}
is concerned with two-dimensional surfaces in a compact 3-manifold.
These surfaces are composed of the unions of `trinions' -
intersections of three `ribbons' - and are scored to divide the
surface into clearly demarcated trinions (fig.\ref{figure1}).

\begin{figure}[!h]
  \begin{center}
    \includegraphics[scale=0.2]{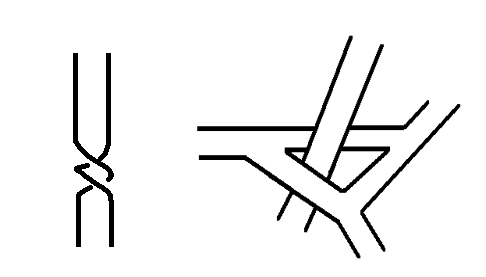}
  \end{center}
\caption{Twists and Braidings} \label{figure2}
\end{figure}

\begin{figure}[!h]
  \begin{center}
    \includegraphics[scale=0.2]{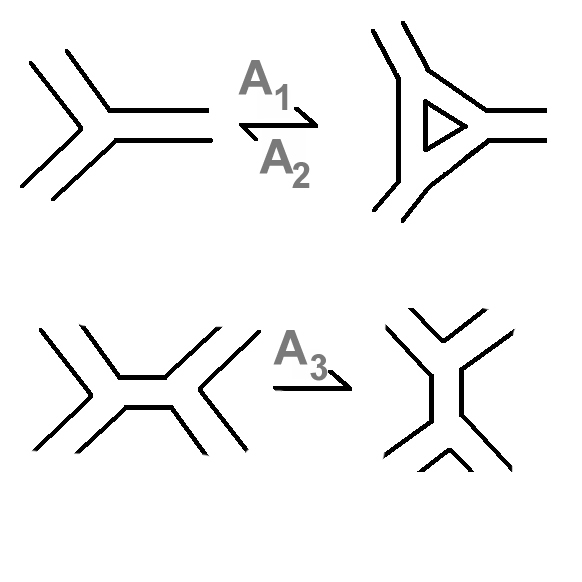}
  \end{center}
\caption{Generators of $\mathcal{A}_{evol}$} \label{figure3}
\end{figure}

 We allow the ribbons to be braided through the punctures in the surface, and we
also allow the ribbons to be twisted by multiples of $2\pi$
(fig.\ref{figure2}). This network evolves under
$\mathcal{A}_{evol}$, the algebra generated by the elements $A_1$,
$A_2$ and $A_3$ (fig.\ref{figure3}).   By viewing the trinions as
nodes, we can consider the manifolds to be graphs.  The theory is
then similar to loop quantum gravity in its structure, though with
some additional allowances for the labelings of the graph.  We also
note that a graph can be changed to a ribbon graph by `framing' the
edges of the graph: turning the one-dimensional edges into
two-dimensional surfaces.

The reduced link of a graph is taken by treating each edge of the
ribbons to be a strand and then excluding unlinked unknotted
strands.  A subsystem is then defined as a section of the graph
where its reduced link does not intersect the rest of the graph.

\begin{figure}[!h]
  \begin{center}
    \includegraphics[scale=0.2]{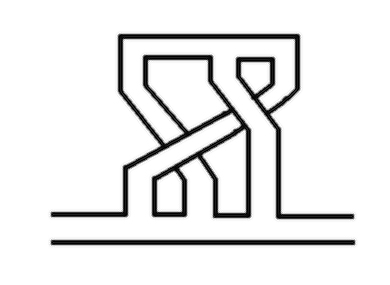}
  \end{center}
\caption{Proposed form of the first generation} \label{figure4}
\end{figure}

The first generation of the standard model is then proposed to be
generated by placing $2\pi$ twists on the strands of the two
crossing capped braid of three ribbons (fig.\ref{figure4}), subject
to the restrictions that all the twistings on a braid must be in the
same direction.

\section{Topology}
In order to properly discuss the idea of a translation we must first
discuss the topology with respect to which the translations shall
occur.  Ribbon networks present difficulties in this regard as there
are several distinct classes of topologies.  We begin by considering
the topology inherent in the idea of neighbours from a graph theory
perspective.

\begin{quote}
\textbf{\textit{The Microlocal Metric Space}}

Consider a ribbon graph $\Gamma$ consisting of $N$ nodes and $M$
ribbons having some braiding and twisting content. We construct a
new metrical space $\tilde{\Gamma}$ as follows: let $X$ be the set
of trinions within the ribbon graph.  We shall take each $x$ within
$X$ as a node in a pseudograph, and construct edges for this
pseudograph in the natural way: by making an edge between two nodes
if their respective trinions share a scored ribbon. This is the
reverse of the framing process that can be used to construct a
ribbon network.

\begin{quote}
\textbf{\textit{The Microlocal Distance Function}}
\\
 Considering the set of all possible paths between two nodes on the
 pseudograph, the distance between the nodes is the minimum number of
 edges in any such path.  This satisfies the four requirements for a
 distance function: that it is positive for any choice of two nodes, that it is  strictly positive for any
 two non-identical nodes, that it is  reflexive and that it satisfies the triangle
 inequality.  This metric is equivalent to the standard metric of
 graph theory.
\end{quote}

Thus, the set $X$ of nodes, along with the microlocal distance
function, create a metric space and, therefore, have a standard
topology $\mathcal{T}_1$ defined by the open balls given by the
microlocal distance function on $X$. $\mathcal{T}_1$ is thus the
induced graph topology of the graph $\Gamma$.
\end{quote}

 The microlocal topology surprisingly contains very little
information about the structure of the ribbon network.  We should
therefore consider topologies that contain information about the
braidings and twists of the ribbons.

\begin{figure}[!h]
  \begin{center}
    \includegraphics[scale=0.1]{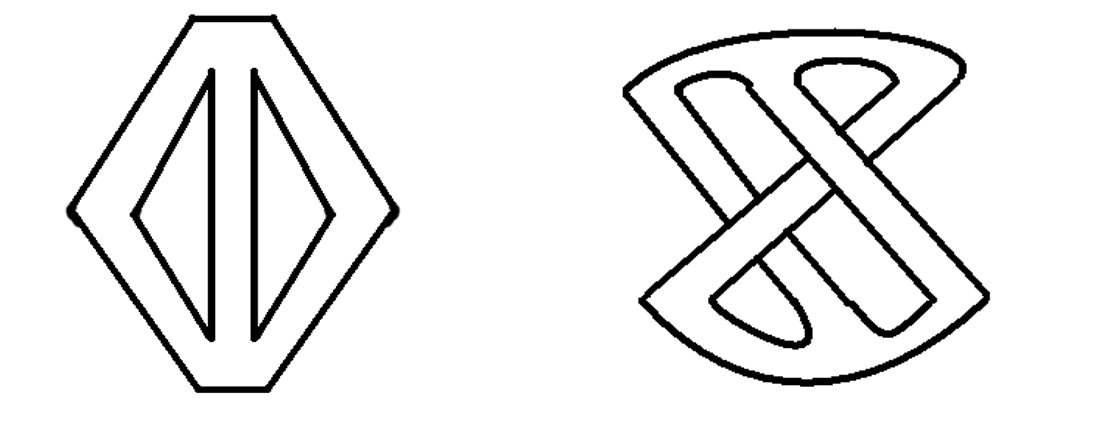}
  \end{center}
\caption{Networks that can be differentiated by $\mathcal{T}_2$}
\label{t2pic}
\end{figure}

\begin{quote}
\textbf{\textit{Ribbon Topology}}\\
 The \textit{Ribbon Topology} is defined to be the topology corresponding to taking the ribbon
network as a bounded two dimensional surface with a Euclidean
metric. The Euclidean metric, together with the bounded space of the
ribbons, then becomes a metric space.  Again, the open balls
generate the topology $\mathcal{T}_2$.  In contrast to
$\mathcal{T}_1$, $\mathcal{T}_2$ is able to differentiate between
graphs like those in figure \ref{t2pic}.
\end{quote}

\begin{figure}[!h]
  \begin{center}
    \includegraphics[scale=0.1]{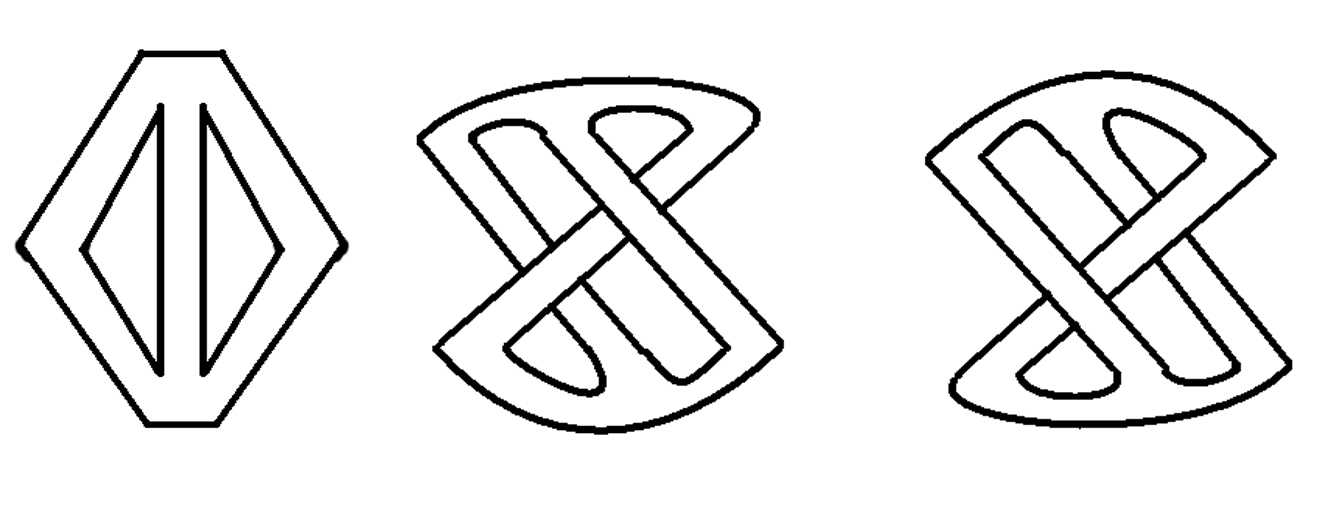}
  \end{center}
\caption{Networks that can be differentiated by $\mathcal{T}_3$}
\label{t3pic}
\end{figure}

\begin{quote}
\textbf{\textit{Braided Ribbon Topology}}\\
 The \textit{Braided Ribbon
Topology} is defined to be any topology of the ribbon network that
includes the braiding and the twisting of the ribbons.  We shall
call this topology $\mathcal{T}_3$.  As the twisting is `invisible'
to anything living directly on the ribbons, this topology has to
appeal to the higher space that the structure is embedded within.
$\mathcal{T}_3$ would then be able to differentiate between graphs
like those in figure \ref{t3pic}. In the same way that we have
referred to the `microlocal' character of objects, we shall refer to
the `braidedlocal' character of objects.
\end{quote}

If we consider $\mathcal{T}_1$, $\mathcal{T}_2$ and $\mathcal{T}_3$
to be topologies on the original network $\Gamma$, we can see that
each successive topology is finer than the last, with
$\mathcal{T}_3$ being the finest.

\section{$\mathcal{A}_{evol}$ and Microlocal Translations}

We shall now demonstrate that there are indeed translations of
braidedlocal structures with respect to the microlocal distance
function. Also, we shall demonstrate that even when using a more
general notion of microlocal distance we can nonetheless demonstrate
situations where braidedlocal structures have undergone a
translation.  These translations are generated by
$\mathcal{A}_{evol}$.

We shall first introduce a series of definitions and then prove a
result using them.

\begin{quote}
\textbf{\textit{Ribbon Connected}}\\
 Two nodes $a$ and $b$ are
\textit{Ribbon Connected} if there exists a sequence of $N+1$ nodes
 $x_n$ such that $x_1 = a$, $x_{N+1} = b$
and for each $n$ the trinion with node $x_n$ and the trinion with
node $x_{n+1}$ share a scored ribbon.  This is equivalent to the
nodes being connected in the graph $\tilde{\Gamma}$.
\end{quote}

\begin{quote}
\textbf{\textit{Connected Ribbon Network}}\\
 A \textit{Connected
Ribbon Network} is a set of nodes $X$, such that all nodes in $X$
are ribbon connected to all other nodes in $X$.
\end{quote}

\begin{quote}
\textbf{\textit{Edge Segments}}\\
 Consider the edges of a ribbon graph
as a metric space $E$ onto itself.  This space is essentially a
collection of 1-d spaces which can be mapped to the unit circle with
the distance between two points being the minimum angle between the
points on the unit circle.  An \textit{Edge Segment} is then any
connected subset of $E$ with a non-empty interior. We consider only
sets of non-empty interior to avoid singleton sets that can produce
difficulties in later considerations.
\end{quote}
\begin{quote}
\textbf{\textit{Edge Connected}}\\
 Two edge segments $a$ and $b$ are
\textit{Edge Connected} if they are connected in the metric space
$E$.
\end{quote}

\subsection{Isolated Substructures}

In order to demonstrate translations within ribbon networks we must
first define a special class of elements within ribbon networks.

\begin{quote}
\textbf{\textit{Isolated Substructure}}\\
 An \textit{Isolated
Substructure} is a ribbon connected set of nodes where a closed
surface can be placed around it with exactly one ribbon intersecting
the surface. We call this ribbon the Isolated Substructure's
``tether''.
\end{quote}

It should be understood that isolated substructures are not the same
 as `subsystems' as defined by \cite{Bilson-Thompson:2006yc}.  This is readily apparent by considering the form of of the reduced link
 of an isolated substructure.

\begin{figure}[!h]
  \begin{center}
    \includegraphics[scale=0.2]{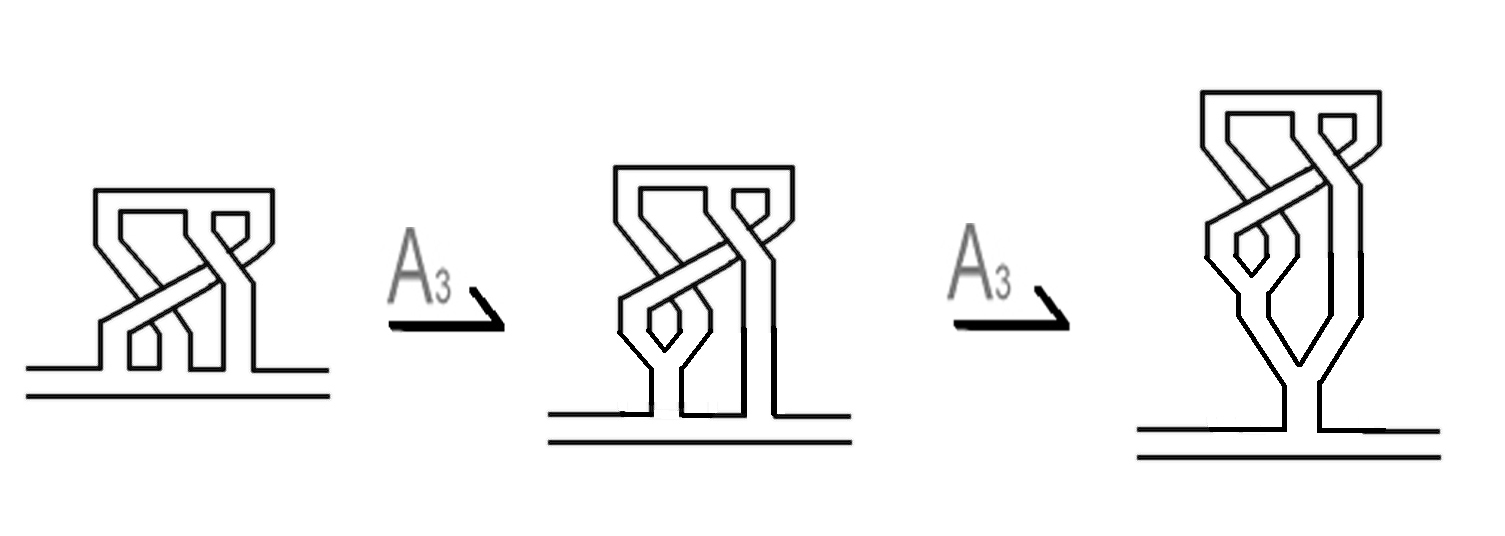}
  \end{center}
\caption{Transforming a capped braid into an isolated substructure}
\label{bundle}
\end{figure}

It is interesting to note that, though the definition of an isolated
substructure appears to be restrictive at first glance, there are a
significant number of structures that can be `packed up' into the
form of an isolated substructure.  For instance, all of the example
definitions of particles from \cite{Bilson-Thompson:2006yc} can be
changed into isolated substructures through the use of exchange
moves from $\mathcal{A}_{evol}$ as shown in figure \ref{bundle}.

\begin{quote}
\textbf{\textit{Replaceable Edge Segments}}\\
 An edge segment is
\textit{Replaceable} if it can be unambiguously replaced by an
isolated substructure's tether. Specifically, a \textit{Replaceable
Edge Segment} cannot be the edge of a node as this would cause four
valent nodes - such nodes are prohibited by our construction.
\end{quote}

These definitions together allow us to consider the dynamics of
isolated substructures under the generators of $\mathcal{A}_{evol}$.
We can consider a graph $\Gamma$ to be composed of a set of isolated
substructures attached to replaceable edge segments of a second
graph $\Lambda$.  As we do not require that all such isolated
substructures be so removed, this procedure can be done without
ambiguity.

\begin{figure}[!h]
  \begin{center}
    \includegraphics[scale=0.3]{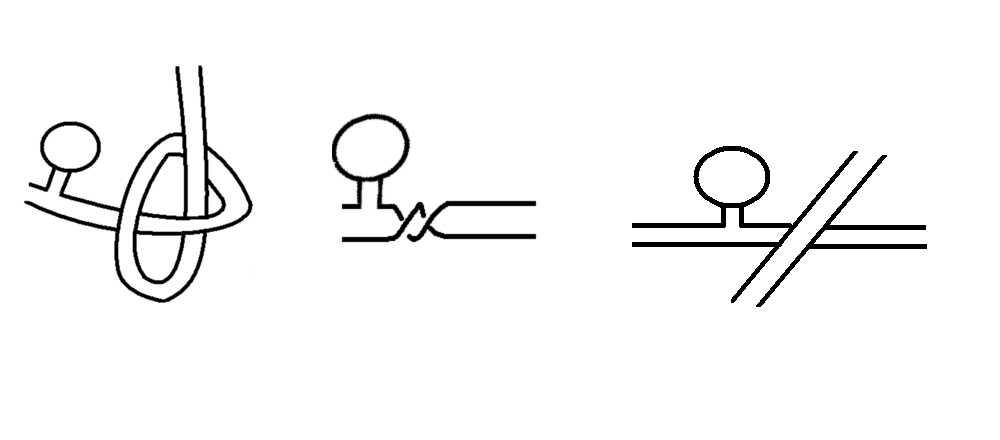}
  \end{center}
\caption{Examples of Terrain} \label{terrain}
\end{figure}

\begin{quote}
\textbf{Theorem}\\
Given a finite closed network $\Gamma$ with two edge connected
replaceable edge segments $a$ and $b$, there exists a sequence of
generators of $\mathcal{A}_{evol}$ such that a graph $\Gamma_a$
 - composed of $\Gamma$ with an isolated substructure $A$ tethered to
$a$ - evolves to $\Gamma_b$, where $\Gamma_b$ is composed of the
same graph $\Gamma$ but with $A$ now tethered to $b$.
\end{quote}
\begin{quote}
\textbf{Proof}\\
We shall proceed by induction on the number of nodes between $a$ and
$b$, say $N$. As $a$ and $b$ are edge connected, the node created by
$A$ being tethered to $a$ is ribbon connected to the two nodes that
are at either side of $b$. We shall label these nodes $x_0$ (for the
node created by $A$ at $a$) through $x_{N+1}$ in such a way that
each $x_{j}$ shares a single ribbon with $x_{j+1}$.  The nodes on
either side of $b$ are then labeled $x_{N}$ and $x_{N+1}$.

Before we perform this induction, we need to show the ability to
move an isolated substructure through intermediate topological
structures that are not composed of nodes.  These are comprised of
three categories: knots, twists and braidings.  Examples of each of
these is shown in figure \ref{terrain}.  As isolated substructures
only have a single connection to the outside network, we can move it
past this `terrain' through the following procedures.
\end{quote}

\begin{quote}
\textit{ For each knot the isolated substructure is pulled through
the knot by stretching out the knot until the substructure can pass
through it. As the substructure is unconnected except through its
tether, this leaves the network unchanged other than the reversal of
the position of the knot and the substructure.}

 \textit{For each
twist (consisting of a rotation by $\pi$), the isolated substructure
can run along the edge of the twist. Alternatively, one can view the
procedure as deforming the network itself by twisting the segment
with the isolated substructure in a manner that undoes the twist on
the one side and create a twist on the other.}

\textit{For each braiding, we deform the ribbon of the braid by
sliding it over the isolated substructure to the other side.  This
is reminiscent of the Reidemeister move of the second kind.}

Our lemma is that we can move an isolated substructure tethered to a
node $R$ so that it is tethered with no intermediate `terrain'
between it and some edge segment $t$ (which is not a component of a
piece of `terrain') that connects the node $R$ to its nearest
neighbours and is edge connected to the edge which the isolated
substructure is tethered to. This is proven by induction on the
number of elements of `terrain' between $R$ and $t$ and the use of
the above prescriptions.  A consequence of this is that the same
method can be used for a node $S$ which has microlocal distance 1 to
$R$.  This ability shall be used heavily in our proof.

Now, returning to the proof, we shall first prove the case of $N=1$.
We apply the above lemma to move the isolated substructure through
any intermediate terrain between $x_0$ and $x_1$, giving us $A$
tethered to a new node $x_0'$ (we shall use primes to denote nodes
that have undergone some change) that is immediately adjacent to
$x_1$ with no intermediate terrain. We then perform an exchange move
from $\mathcal{A}_{evol}$ on the node $x_0'$ and $x_1$ to move
$x_0'$ onto the edge on the other side of $x_1'$. We can then again
use the above lemma (in its more general case) to move $x_0'$ to its
final resting place at $b$.

Now we shall assume that the case of $N-1$ nodes is correct and
prove the case of $N$ nodes.  The prescription for this is analogous
to the $N=1$ case.  Given that there is a method for moving past
$N-1$ nodes (by inductive hypothesis), we shall use that method to
change the situation to a single intermediate node, and then invoke
the method of the $N=1$ case to bypass the final node.

The preceding gives the inductive argument and completes the proof.
\end{quote}

To demonstrate translations we will need a further tool.  We
therefore consider also the following lemma:
\begin{quote}
\textbf{Lemma} \textit{Translations Through an Isolated
Substructures}\\
 Given an isolated substructure $A$ that has been
moved to the edge of the tether of another substructure $B$, it is
possible to translate $A$ to the opposite edge of the tether of $B$.
\end{quote}
\begin{quote}
\textbf{Proof}\\
Due to the above theorem, it only remains to show that the two edges
of the tether of an isolated substructure are edge connected.
Proceeding by contradiction, we assume that they are not edge
connected.  As we see that an edge of the network enters the
isolated substructure and does not exit, there must be some terminus
of the edge within the isolated substructure. However, such a
situation is impossible, as the edges of a ribbon network must form
closed links or terminate at some boundary (which we have not
introduced into the theory of ribbon networks).  We therefore have a
contradiction.  Thusly we see that if an isolated substructure can
be moved to the edge of a tether, by the above theorem, it can be
moved to the other edge.
\end{quote}

\subsection{Microlocal Translations}

\begin{figure}[!h]
  \begin{center}
  \subfloat[]{\label{trans1}\includegraphics[scale=0.3]{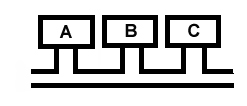}}
  \subfloat[]{\label{trans2}\includegraphics[scale=0.3]{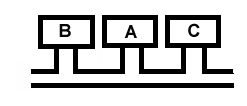}}
\end{center}
  \caption{Microlocal Translations}
  \label{mtrans}
\end{figure}

The application of the theorem is straightforward and results in the
ability to demonstrate translations under the microlocal distance
function.  For instance, it is possible to construct a sequence of
moves of $\mathcal{A}_{evol}$ such that figure \ref{trans1} evolves
to figure \ref{trans2}.  Under the microlocal distance function, the
isolated substructure $A$ is now less distant from the isolated
substructure $C$ (measuring the distance between substructures from
the node at which they are tethered).

Even applying a more restrictive definition of distances, we can
demonstrate translations in some form. Consider the following
definition of closeness:

\begin{quote}
\textbf{\textit{$\alpha$-closer}}\\
 An isolated substructure $A$ issaid to be \textit{$\alpha$-closer} to an isolated substructure $B$
than it is to another substructure $C$, if for all paths along the
ribbons of the network, leaving the node at which $A$ is tethered
and intersecting the node at which $C$ is tethered the path
intersects the node at which $B$ is tethered.
\end{quote}

\begin{figure}[!h]
  \begin{center}
  \subfloat[]{\label{atran1}\includegraphics[scale=0.3]{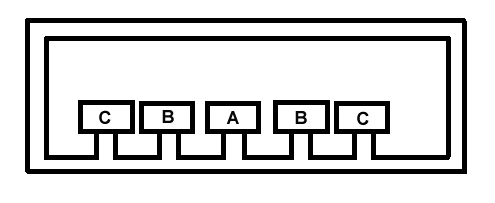}}
  \subfloat[]{\label{atran2}\includegraphics[scale=0.3]{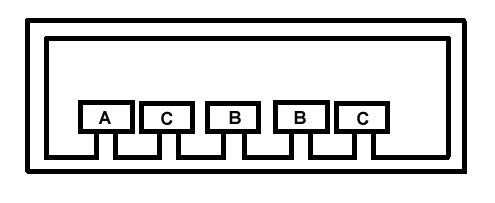}}
\end{center}
  \caption{$\alpha$-closer Translations}
  \label{atrans}
\end{figure}

It is difficult to find such a situation where translations with
respect to this definition can be demonstrated clearly. However, if
we expand the definition slightly to allow us to consider isolated
substructures with identical structure to be treated equally, we can
show that it is possible to evoke a translation of a certain form.
Specifically, it is possible to take a situation where a
substructure $A$ is $\alpha$-closer to substructures of type $B$
than to those of type $C$ and to apply a series of moves of
$\mathcal{A}_{evol}$ such that the reverse is true afterwards. For
instance, consider figure \ref{atran1} and figure \ref{atran2}. Thus
we see that we have translations even under stringent requirements,
thereby concluding our result.

\section{Macro, Micro and Braided Locality}
\label{nogo}

The braidedlocal structure of the braid network is characterized by
the reduced link of the structure. The reduced link of a braid
network can be shown to be invariant under the generators of
$\mathcal{A}_{evol}$, by applying the definition of the reduced link
to the graphical representations of the generators.  As a result, it
is clear that the braidedlocal content of a braid network is
invariant under the generators of $\mathcal{A}_{evol}$. This
invariance is a double edged sword.  On one hand, it allows us to
assign some meaning to these invariant structures, as was done by
the authors of \cite{Bilson-Thompson:2006yc}.  On the other hand, it
means that there is no way that these microlocal moves can provide
any form of dynamics in this content.  As a result, I suggest that,
to construct a theory of quantum gravity containing particle physics
from a ribbon network, it is necessary to consider the existence of
a second evolution algebra, which I shall call
$\mathcal{A}_{braid}$.

In \cite{Markopoulou:2007ha} and \cite{Markopoulou:2006qh}, the
concept of macrolocality in networks is put forward as the locality
derived from the classical metric that would arise for a network
with a space-time as its classical limit.  In
\cite{Markopoulou:2007ha}, the authors then remind us that there is
no need for macrolocality to be coincidental to microlocality.  It
seems to be a consequence of the ideas of
\cite{Bilson-Thompson:2006yc} to suggest that, though microlocality
and macrolocality are not necessarily coincidental, the braidedlocal
content - the invariants that we associate with particles - should
be part of the bridge of the gap between the two. I therefore
suggest that $\mathcal{A}_{braid}$ could be the bridge between
microlocality and macrolocality.

For future consideration, I outline some general possibilities of
$\mathcal{A}_{braid}$. Regardless, it should be noted that any such
algebra that could provide macrolocal dynamics, particle
interactions included, would need to alter the reduced link of the
network if the identifications in \cite{Bilson-Thompson:2006yc} are
to be considered seriously.

\begin{quote}
\textbf{\textit{Nearly Microlocal Algebra}}\\
 A candidate based upon
the assumption that any move within the second evolution algebra
should be as close to being microlocal as possible is called a
\textit{Nearly Microlocal Algebra}.  This could be completed by
introducing moves involving next to nearest neighbor nodes.  This
suggestion corresponds to the idea that there is a degree of
coincidence between microlocality and macrolocality (again, we
should remember that such a coincidence is not
needed).\cite{Markopoulou:2007ha}
\end{quote}

\begin{quote}
\textbf{\textit{Braid Algebra}}\\
 An algebra based upon moves that
alter the braiding content of the network in ways that are roughly
equivalent to elements of the standard braid group is referred to as
a \textit{Braid Algebra}. Also, it can contain moves that allow the
composition of multiple braided isolated substructures.
\end{quote}

\begin{quote}
\textbf{\textit{Anti-Microlocal Algebra}}\\
 An algebra premised upon
the idea that microlocality should be dual or completely unrelated
to macrolocality is called an \textit{Anti-Microlocal Algebra}. Such
an algebra can be constructed from a set of moves that act upon the
reduced links of a graph.  Such moves could be realized through the
following algorithm:
\begin{quote}
Take the reduced link of the graph $\Gamma$ and apply a move that
composes or interacts parts of the reduced link (whether through
cutting and repairing links, or through allowing links that
correspond in some manner to annihilate each other).  Then take the
new reduced link and equate it with a superposition of all graphs
$\Gamma_x\prime$.  That any such graph $\Gamma_x\prime$ should exist
should be provable by a generalization of the theorem that allows
the construction of a closed braid that corresponds to any link.
\cite{braids}
\end{quote}
\end{quote}

It is possible that a stronger candidate would draw upon multiple
such programs.

\section{Conclusion}

The above results give rise to several key points.  First, the
results are restricted to isolate substructures, without which it is
impossible to bypass the terrain within the network.  Second, the
definitions in the previous section may not necessarily apply if
labels are introduced to the network.  Despite these restrictions,
the result remains promising and integral to attempts to attempts to
develop Ribbon networks into a theory of quantum gravity with
matter.  The primary candidates for the fundamental particles within
such a theory are all examples of systems that can be made into
isolated substructures.  Indeed the form of the fundamental
particles was the motivation for demonstrating translations.

The demonstration of these translations provides great promise in
further developing this model into a theory that involves particle
dynamics.  However several key obstacles remain. As discussed in
section \ref{nogo}, without adding more structure, in the form of a
a second evolution algebra (or at the least, expanding the original
evolution algebra), it is impossible to have any particle
interactions.  Indeed, even the case that one might expect to be
easiest to demonstrate - that of particle and anti-particle
annihilating one another - is impossible without some modification.
 Developing candidates for $\mathcal{A}_{braid}$ remains the subject
of ongoing work.

\section{Acknowledgements}
I would like to thank Lee Smolin, Fotini Markopoulou-Kalamara,
Sundance Bilson-Thompson and Louis Kauffman for many helpful
discussions related to the material presented herein.

Research at Perimeter Institute for Theoretical Physics is supported
in part by the Government of Canada through NSERC and by the
Province of Ontario through MRI.

\end{document}